\documentclass[prl,aps,showpacs,amsmath,amssymb,twocolumn]{revtex4-1}
\usepackage{graphicx}
\usepackage{dcolumn}
\usepackage{bm}
\usepackage{hyperref}

\def\mnras{MNRAS}
\def\apj{ApJ}

\begin{document}
\title{Satellite Test of the Equivalence Principle\\as a Probe of Modified Newtonian Dynamics}

\author{Jonas P. Pereira$^{1}$, James M. Overduin$^{1}$ and Alexander J. Poyneer$^{1}$} 
\email{e-mail to: joverduin@towson.edu}
\affiliation{$^{1}$Department of Physics, Astronomy and Geosciences, Towson University, 8000 York Road, Towson, Maryland 21252, USA}

\date{\today}

\begin{abstract}
The proposed Satellite Test of the Equivalence Principle (STEP) will detect possible violations of the Weak Equivalence Principle by measuring relative accelerations between test masses of different composition with a precision of one part in $10^{18}$.
A serendipitous byproduct of the experimental design is that the \textit{absolute} or common-mode acceleration of the test masses is also measured to high precision as they oscillate along a common axis under the influence of restoring forces produced by the position sensor currents, which in  drag-free mode lead to Newtonian accelerations as small as $10^{-14} g$. This is deep inside the low-acceleration regime where Modified Newtonian Dynamics (MOND) diverges strongly from the Newtonian limit of General Relativity.
We show that MOND theories (including those based on the widely-used ``$n$-family'' of interpolating functions as well as the covariant Tensor-Vector-Scalar formulation) predict an easily detectable increase in the frequency of oscillations of the STEP test masses if the Strong Equivalence Principle holds. If it does not hold, MOND predicts a cumulative increase in oscillation amplitude which is also detectable. STEP thus provides a new and potentially decisive test of Newton's law of inertia, as well as the equivalence principle in both its strong and weak forms.
\end{abstract}

\maketitle

\section{Introduction}

Newtonian physics (including the Newtonian limit of General Relativity) fails badly on the scales of galaxies and galaxy clusters, where it predicts much smaller velocities than are actually observed. The discrepancy is widely attributed to gravitational attraction from dark matter (DM), but the implied DM density is five times that of all known standard-model (baryonic) matter in the Universe, and no DM candidates have yet been detected either directly (in laboratory experiments) or indirectly (by means of decays or annihilations that could contribute to astrophysical backgrounds), despite decades of searching \cite{overduin2008light}. An alternative, called Modified Newtonian Dynamics (MOND), is to modify either Newton's law of gravitation or the law of inertia for accelerations less than $a_0\approx c H_0\approx 10^{-10}$~ms$^{-2}$ \cite{1983ApJ...270..365M,sanders2002modified,2012LRR....15...10F}. The gravitational implementation of MOND implies a modification of General Relativity in principle \cite{2004PhRvD..70h3509B}, while the inertial one entails modifying the standard kinetic term of Lagrangian mechanics, which would affect not only gravitational phenomena but those generated by all the other fundamental interactions as well \cite{2011arXiv1111.1611M}.

Either way, it is plainly desirable to test MOND away from the astrophysical context in which it was conceived. We consider here STEP (the Satellite Test of the Equivalence Principle), in which pairs of test masses with superconducting coatings orbit the Earth in free fall, their motions along a common axis monitored to high precision by SQUID (superconducting quantum interference device) magnetometers \cite{1978AcAau...5...27W}. STEP is designed to detect \textit{differences} in acceleration between test masses of different composition, which would violate the Weak Equivalence Principle (WEP). MOND is known to satisfy the WEP \cite{1984ApJ...286....7B}, so such a test might not appear relevant at first glance. However, STEP is designed so that differential and common accelerations are measured by separate SQUIDs, with a common-mode sensitivity of $10^{-18} g$ in drag-free mode \cite{2001CQGra..18.2475M}. Furthermore, the SQUID circuits exert small restoring forces on the test masses, causing them to oscillate with periods of order 1000~s and local Newtonian accelerations no larger than $10^{-13}$~ms$^{-2}$, well below $a_0$. The possibility thus arises of using STEP simply as a way to check whether or not a test mass on the end of a spring is governed by Newtonian or MONDian dynamics in the low-acceleration regime. Full technical details on the experiment are found in \cite{1978AcAau...5...27W,2001CQGra..18.2475M}, and its current status and scientific motivation have recently been discussed in \cite{2012CQGra..29r4012O}. Our main conclusion in this \textit{Letter} is that STEP has the unique capability to provide a new and potentially decisive test, not only of the WEP, but of the SEP, and of Newton's law of inertia itself. The revival of the STEP mission (currently dormant due to lack of funding) should therefore be a top priority in fundamental physics.

Although MOND fulfills the WEP, the Strong Equivalence Principle (SEP) is almost certainly violated if the modifications involve only the gravitational sector \cite{1984ApJ...286....7B}. Within the inertial implementation of MOND the status of the SEP is still an open issue \cite{2009MNRAS.399..474M,2011arXiv1111.1611M}. 
Thus test-mass behavior may be affected by STEP's acceleration relative to an inertial frame. Since this ``external acceleration'' is large compared to $a_0$, the dynamics may be close to Newtonian, even if ``internal'' accelerations are much smaller than $a_0$. This is referred to as the External Field Effect (EFE), and is a major reason why existing proposals for laboratory tests of MOND have not been regarded as conclusive to date \cite{2007PhRvL..98o0801G,2007MNRAS.377L..79F}.

An important step was taken by Ignatiev \cite{2007PhRvL..98j1101I, 2008PhRvD..77j2001I}, who showed that EFEs could be minimized on Earth during specific periods of the year. An improvement on this idea was suggested in Ref.~\cite{2009A&A...503L...1D}. 
A simpler approach has recently been proposed by Das and Patitsas \cite{2013PhRvD..87j7101D}, who made use of free falling laboratories and assumed the validity of the SEP. Under these conditions, they showed that MONDian predictions could easily be differentiated from their Newtonian counterparts for some specific experiments. In what follows, we apply a similar analysis to STEP, initially assuming that the SEP is valid. However, given that the SEP should \textit{not} generally be assumed in the context of MOND, we go further and allow for the possibility of SEP violation, explicitly including EFEs in our calculations. This requires us to treat MONDian effects in noninertial coordinate systems in some detail, for the first time as far as we are aware.

\section{MOND with SEP}

We assume to begin with that the SEP is valid, so that the outcome of any experiment performed within the ``local'' (internal) STEP frame is independent of the acceleration and space and time location of the spacecraft itself relative to an external inertial frame \cite{1984ApJ...286....7B,2014LRR....17....4W}\footnote{In this work we analyze a non-gravitational experiment, which means that the SEP is formally equivalent to the Einstein Equivalence Principle (EEP) \cite{2014LRR....17....4W}.}. Though the STEP payload is complex, the kinematics of the test masses is simple. The presence of magnetic fields due to the currents in the superconducting circuits leads to spring-like restoring forces with typical periods of approximately $1000$~s and nominal amplitudes of $10^{-10}-10^{-8}$~m relative to the spacecraft during normal drag-free operation \cite{2001CQGra..18.2475M} \footnote{STEP is ingeniously designed to eliminate gravity-gradient forces (Earth gradients produce a signal at a known frequency, which is used to null out spacecraft gradients through small adjustments of the test masses \cite{2001CQGra..18.2475M}). The gravitational field of the spacecraft itself also becomes important in principle when one is measuring absolute, as opposed to relative accelerations. This field is determined by the distribution of mass, which can always be designed such that the resultant force on  given test masses is perpendicular to their direction of motion, which is thus counteracted by the normal constraint forces of the magnetic bearings.}.
Thus the maximum Newtonian acceleration of the test masses with respect to the STEP frame is of order $10^{-15}-10^{-13}$~ms$^{-2}$, far below the MOND scale $a_0$.
In its inertial formulation, MOND specifies the force $\vec{F}$ acting on a body of inertial mass $m_i$ as
\begin{equation}
\vec{F}=m_i \mu\left(\frac{|\vec{a}|}{a_0}\right)\vec{a}\label{ForceMond},
\end{equation}
where $\mu(x)$ is the interpolating function, whose precise form is not yet known, but which must go over to $\mu(x)\approx 1$ when $x\gg 1$ (Newtonian regime) and $\mu(x)\approx x$ when $x\ll 1$ (deep MOND regime). 

Given that the accelerations of the test bodies with respect to a local frame in  drag-free mode are at least three orders of magnitude below the MOND scale, we are in the deep MOND regime and  $\mu(x)\approx x$ regardless of the specific choice of MOND theory. We orient our internal frame such that one axis ($\hat{z}$, say) coincides with the direction of motion of a given pair of test masses and identify its origin with the equilibrium point of the effective restoring force. Eq.~(\ref{ForceMond}) then gives for $z>0$
\begin{equation}
\ddot{\bar{z}}+ \sqrt{\frac{a_0}{z_0}}\omega\sqrt{\bar{z}}=0 \label{MONDeq},
\end{equation}
where overdots denote time derivatives, $\bar{z}\doteq z/z_0$ is normalized displacement, $z_0$ is the amplitude of the oscillations, and $\omega$ is their Newtonian frequency (defined in terms of the inertial mass). We choose initial conditions at the point of maximum displacement such that $\bar{z}(0)=1$ and $\dot{\bar{z}}(0)=0$. From symmetry, it is sufficient to analyze the case $z>0$. Restoring forces guarantee that the motion is still oscillatory and its amplitude remains fixed.
Acceleration is larger than in the Newtonian case [$\bar{z}_{N}(t)=\cos(\omega t)$] due to the asymptotic property of $\mu(x)$, leading to shorter oscillation times.

\begin{figure}[t!]
    \centering
    \includegraphics[width=\columnwidth]{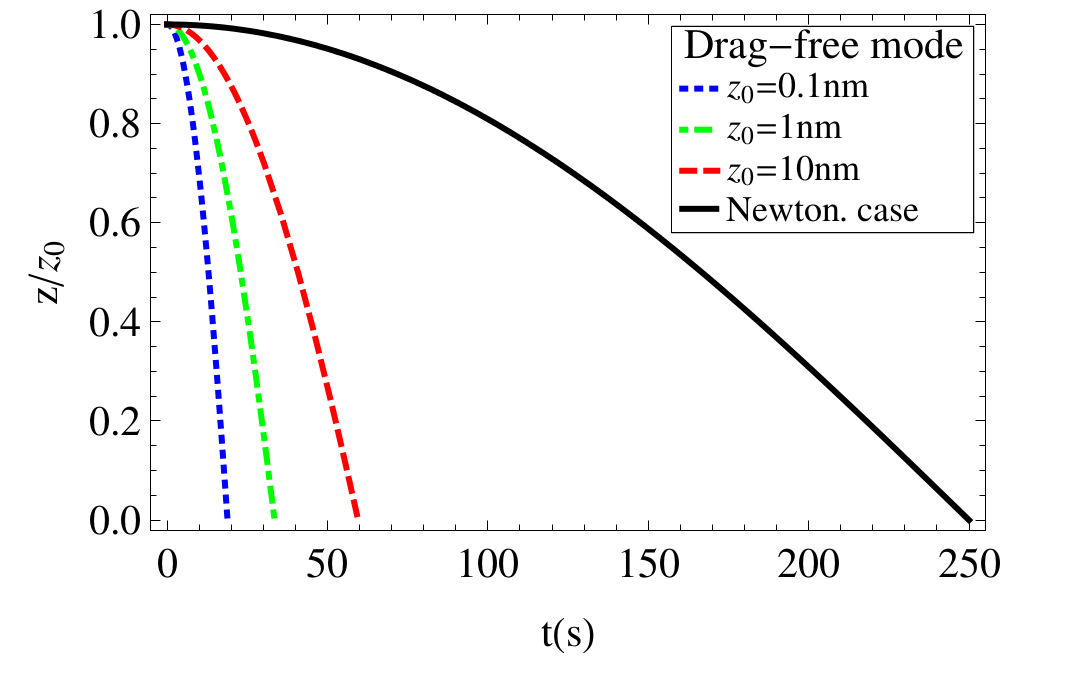}
    \caption{(color online). Numerical integration of Eq.~(\ref{MONDeq}) for different amplitudes of oscillation of the STEP test masses (initial conditions), assuming that the SEP is valid within MOND and that drag-free conditions are maintained. Oscillation frequency depends on amplitude (but remains fixed for a given choice of initial condition), and differs strongly from its Newtonian counterpart [$\bar{z}_{N}(t)=\cos(\omega t)$] in every case.}
    \label{fig:deep_mond}
\end{figure}

Eq.~(\ref{MONDeq}) can be solved numerically or analytically in terms of hypergeometric functions, with results as shown in Fig.~\ref{fig:deep_mond}.
Note that the amplitude of motion influences the period of the system (from the symmetry of the problem it is simply $4t_{0}$, where $\bar{z}(t_0)=0$). Thus the oscillation frequency for a given amplitude is a natural choice of a physical observable. We shall come back to this issue later in this section.

We now consider the scenario in which the SEP is still satisfied, but in which amplitudes can take arbitrarily large values up to the hard limit (of order mm) imposed by the experimental design \footnote{For definiteness, we assume here that the effective  potential continues to have a harmonic form in non drag-free mode.}. In this case, accelerations will exceed the MOND scale $a_0$, so that Eq.~(\ref{MONDeq}) no longer holds. To make predictions, it is then necessary to choose a specific interpolating function $\mu(x)$. Historically, spiral galaxy rotation curves have been well fit with the so-called ``simple $\mu$-function'' $\mu_1(x)=x/(1+x)$ \cite{2012LRR....15...10F}. Another common choice, the ``standard interpolating function'' $\mu_2(x)=x/(1+x^2)^{1/2}$, is more compatible with solar system data \cite{2006MNRAS.371..626S}. Both $\mu_1(x)$ and $\mu_2(x)$ belong to an ``$n$-family'' of functions $\mu_n(x)=x/(1+x^n)^{1/n}$ with $n\geq 1$, and observations on both galactic and solar-system scales can be fit with combinations of the cases $n=1$ and $n=2$ \cite{2012LRR....15...10F}. Such functions are however disfavored by recent Cassini data, which prefer $n\gtrsim 3$ \cite{2009MNRAS.399..474M,2014PhRvD..89j2002H}. An alternative is the interpolating function derived by Bekenstein in the context of what is so far the only fully covariant gravitational formulation of MOND, the Tensor-Vector-Scalar (TeVeS) theory,  $\mu(x)_{TVS}=(\sqrt[]{1+4x}-1)/(\sqrt[]{1+4x}+1)$ \cite{2004PhRvD..70h3509B, 2014PhRvD..89j2002H}.
Strong observational constraints on this version of MOND, as well as the more phenomenological $n$-family, have now been reported in Ref.~\cite{2016MNRAS.455..449H}.

It should be stressed that all these constraints have assumed a gravitational formulation of MOND, and do not necessarily apply in the context of MOND as modified inertia. For a comprehensive test of the MOND hypothesis, it is important to compare the predictions of both formulations with experimental data using the same interpolating functions. 
Thus, for specificity, we work in what follows with $\mu_2$ and $\mu_{TVS}$ \footnote{For $\mu_{TVS}$ this is an extrapolation from the original range of validity [$\ll 1$] \cite{2004PhRvD..70h3509B}  that serves as a reference function for data analysis.}. From Eq.~(\ref{ForceMond}) one can show that for $z>0$ the physically relevant equation of motion now for $\mu_2$ is
\begin{equation}
\ddot{\bar{z}}+ \frac{1}{\sqrt{2}}\sqrt{\bar{z}^2\omega^4+ \bar{z}\omega^2\sqrt{\omega^4\bar{z}^2 + \frac{4a_0^2}{z_0^2}}}=0\label{mu2eq},
\end{equation}
while for $\mu_{TVS}$ one gets
\begin{equation}
\ddot{\bar{z}}+ \omega^2\bar{z}+ \sqrt{\frac{a_0}{z_0}}\omega\sqrt{\bar{z}}=0\label{muBek}.
\end{equation}
\begin{figure}[t!]
    \centering
    \includegraphics[width=\columnwidth]{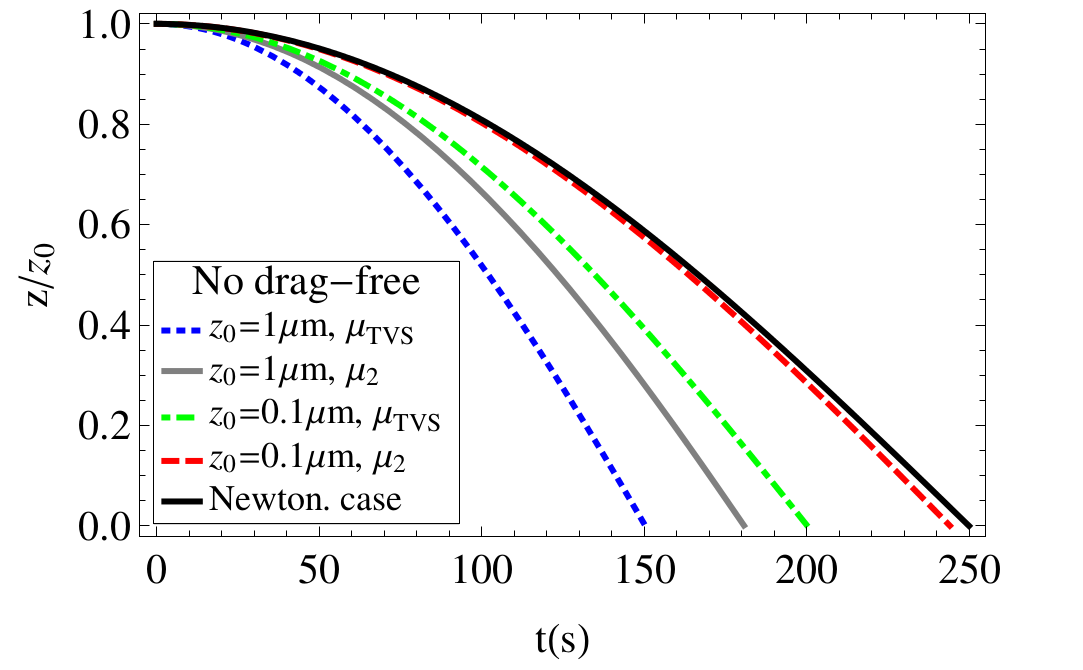}
    \caption{(color online). Numerical integration of Eqs.~(\ref{mu2eq}) and (\ref{muBek}), again assuming that the SEP is valid within MOND (but no longer assuming drag-free conditions). The differences between $\mu_2$ [solid(gray)/dashed] and $\mu_{TVS}$ (dotted/dot-dashed) are detectable for all amplitudes up to order $\sim$mm.}
    \label{fig:no_drag_free}
\end{figure}
%
Numerical integration of Eqs.~(\ref{mu2eq}) and (\ref{muBek}) leads to the results shown in Fig.~\ref{fig:no_drag_free}. Note that $\mu_2$ converges more quickly than $\mu_{TVS}$ to the Newtonian regime, so that they should be easily distinguishable. As expected, the closer to the Newtonian regime (higher amplitudes) the smaller the magnitude of MONDian effects. In particular, for $\mu_2$ with $z_0\simeq$~1~mm it can be verified numerically that the oscillation period is reduced by about $10^{-3}$~s relative to Newtonian expectations. This would however still be readily observable due to cumulative effects from STEP's uninterrupted $10^{6}$ s of data collection \cite{2001CQGra..18.2475M}.

\begin{figure}[t!]
    \centering
    \includegraphics[width=\columnwidth]{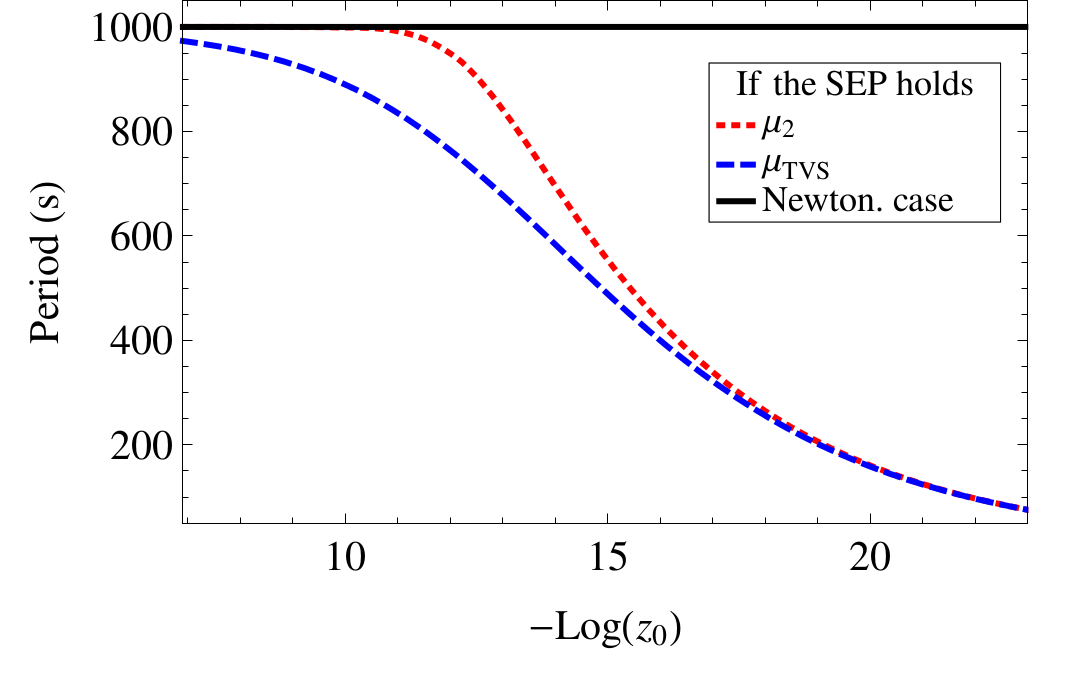}
    \caption{(color online). Period dependence on the amplitude (initial conditions) for $\mu_2$ and $\mu_{TVS}$, assuming no SEP violation. Both interpolating functions agree for small $z_0$ since there accelerations are much smaller than $a_0$, so that both converge to the same asymptotic (deep MOND) limit.}
    \label{fig:period_amplitude}
\end{figure}

We now come back to the issue of the dependence of the frequency of the system on its amplitude (initial condition) when the SEP holds. This is important since periods of oscillation are easy-to-measure observables in this context. (Accelerations near the turning points are also good physical observables, since there $\ddot{z}-\ddot{z}_N\gg 10^{-18} g$ for $\mu_2$ and $\mu_{TVS}$.)
Fig.~\ref{fig:period_amplitude} summarizes this relationship for the whole range of amplitudes STEP may have with respect to $\mu_2$ and $\mu_{TVS}$.
As already indicated by Figs.~\ref{fig:deep_mond} and \ref{fig:no_drag_free}, $\mu_2$ leads to very different predictions when compared to $\mu_{TVS}$, and for small amplitudes the MONDian frequencies are very different from their Newtonian counterparts.

\section{MOND without SEP}

SEP violation in the present context means that the internal dynamics of the system (i.e., the springs and test masses) may be influenced by external properties of the laboratory (i.e., the spacecraft), such as its acceleration with respect to an inertial frame \cite{2014LRR....17....4W}. The first step towards determining MOND's predictions under these circumstances is to derive the equations of motion in noninertial frames. This point is yet unsettled \cite{2007PhRvL..98j1101I} and thus what one could do is to conceive models for them. In this regard, let us simply assume that the vectorial sum of accelerations is valid within MOND.

Consider three reference systems, $S_{in}$, $S'$ and $S$, such that $S_{in}$ is an inertial system, $S'$ only translates with respect to $S_{in}$ and $S$ has a coincident origin with $S'$ and rotates with respect to it with angular velocity $\vec{\Omega}$. The total acceleration of a test particle of inertial mass $m_i$ with respect to $S_{in}$ can then be written as $\vec{a}_{in}=\vec{a}'+\vec{a}+\vec{b}$, where $\vec{a}$ is its acceleration relative to $S$, $\vec{a}'$ is the acceleration of $S'$ with respect to $S_{in}$ and $\vec{b}= \dot{\vec{\Omega}}\times \vec{r}+ 2\vec{\Omega}\times \dot{\vec{r}}+ \vec{\Omega}\times (\vec{\Omega}\times \vec{r})$ takes into account all acceleration terms related to the rotation of $S$ (see, e.g., Ref.~\cite{1969mech.book.....L}; $\vec{r}$ is the radius vector from $S$ to the test particle and $\dot{\vec{r}}$ its velocity.) Multiplying $\vec{a}_{in}$ by $m_i\mu(a_{in}/a_0)$, with $a_{in}\doteq |\vec{a}_{in}|$, leads with the help of Eq.~(\ref{ForceMond}) to the relationship between the acceleration of a test particle relative to a noninertial frame and the forces $\vec{F}$ present there. When applied to STEP, part of $\vec{F}/m_i$ is related to gravitational field of Earth. Assuming that the test particles fall at the same rate as the setup allows us to identify such terms with the motion of the STEP center of mass (c.m).
Thus $\vec{F}/m_i=\vec{F}_{c.m}/M+ \vec{F}_{res}/m_i + \vec{N}/m_i$, $\vec{F}_{res}$ is the effective restoring force on the test particles discussed before, where $M$ is the total mass of the payload and $\vec{N}$ is the sum of normal forces such that the test particles remain along a given axis. Applying the relationship of forces and acceleration discussed above to the motion of the STEP center of mass with respect to $S$ leads to ($y\doteq |\vec{a}'+\vec{a}_{c.m}+\vec{b}_{c.m}|/a_0$)
\begin{equation}
\frac{\vec{F}_{res}}{m_i}+ \frac{\vec{N}}{m_i}\approx \mu(y)\delta \vec{a}_l + y\frac{\partial\mu(y)}{\partial y}\frac{\delta \vec{a}_l\cdot \vec{a}_{c.m}}{ |\vec{a}_{c.m}|^2}\vec{a}_{c.m},\label{FSprime}
\end{equation}
where $\delta \vec{a}_l \doteq \ddot{\vec{r}}_l+2\vec{\Omega}\times \dot{\vec{r}}_l$, $\vec{r}_l$ is the position of the test particle as measured in a local frame freely falling with STEP, $S_l$, and $y\approx |\vec{a}_{c.m}|/a_0$ (because when the payload freely falls, $|\vec{a}_{c.m}|\simeq 10$  ms$^{-2}$, $|\vec{a}'|\simeq |\vec{a}_{c.m}|/1000$ and $|\vec{b}_{c.m}|\simeq |\vec{a}_{c.m}|/10$, so that $y\gg 1$), all related to the acceleration of the center of mass of STEP. Following our previous definition, we take the dynamics of the test particle of interest in the $z$-direction. From the consideration that $\vec{N}$ must exactly cancel out all the perpendicular terms to $\hat{z}$ in Eq.~(\ref{FSprime}) and averaging external quantities over a full orbit (since the only physically meaningful analyses should come from cumulative effects after of multiple orbits \cite{2001CQGra..18.2475M}), we arrive at 
\begin{equation}
\ddot{\bar{z}}\approx -\frac{\omega^2 \bar{z}}{\mu(y) + \frac{y}{2}\frac{\partial\mu(y)}{\partial y} }
\label{effeqmotion},
\end{equation}
since, due to the oscillatory nature of $\vec{a}_{c.m}$ with respect to $S_l$, $\langle(a_{c.m}^{\parallel})^2/|\vec{a}_{c.m}|^2\rangle=1/2$, while $\langle a_{c.m}^{\parallel}a_{c.m}^{\perp}\rangle=0$, with $a_{c.m}^{\parallel}\doteq \vec{a}_{c.m}\cdot \hat{z}$ and $\vec{a}_{c.m}^{\perp}=\vec{a}_{c.m}-a_{c.m}^{\parallel}\hat{z}$. (Here $\langle a_{c.m}^{\parallel}a_{c.m}^{\perp}\rangle=0$, which is is physically equivalent to neglecting Coriolis accelerations when compared to the restoring ones, in agreement with $\Omega\ll \omega$.)
Numerical integration confirms that non-averaged analyses of Eq.~(\ref{FSprime}) for the $\hat{z}$-direction converge to those ones based on Eq.~(\ref{effeqmotion}) after only 2-3 orbits, thus justifying the latter. 

Over the course of multiple orbits, one can constrain the denominator of Eq.~(\ref{effeqmotion}), and thus the function $\mu(y)$, by means of precise distance measurements. Let us first analyze this for the $n$-family of interpolating functions. Since $y\gg 1$ (the spacecraft is in free fall, which corresponds to $y\simeq 10^{11}$), we have $\mu_n(y)=1-1/(ny^n)$. Solving Eq.~(\ref{effeqmotion}), we find that the MONDian solution differs from the Newtonian one by $\sin(\omega t)(n-2) \omega t/(4ny^n)$. Thus, assuming a continuous observation period of order days \cite{2001CQGra..18.2475M}, cumulative effects give rise to a physical difference between MONDian and classical amplitudes of $10^{3}z_0|n-2| /(ny^n)$. (Accelerations are not good physical observables in this context exactly due to their lack of cumulative effects.) Recalling that STEP's precision for position measurements associated with periods of the order of 1000~s and acceleration sensitivity of $10^{-18}$~g is of approximately $10^{-13}$~m \cite{2001CQGra..18.2475M}\footnote{The design of STEP's SQUID circuits leads to a trade-off between position sensitivity ($\Delta z$) and oscillation frequency: for given inductances and couplings, it is such that $\Delta z\propto 1/I$ and $\omega \propto I$, where $I$ is the current \cite{1978AcAau...5...27W}. This automatically results in a trade-off between acceleration precision $\Delta a$ and position sensitivity, since $\Delta a =\omega^2\Delta z \propto I$}, we conclude that detectable amplitude changes are possible when $z_0|n-2| /(ny^n)\gtrsim 10^{-16}$. For $z_0<10^{-4}$~m, there is no $n>1$ that fulfills the above-mentioned inequality. Therefore, STEP is only able to constrain interpolating functions associated with a combination of $n=1$ and $n=2$ if $z_0>10^{-4}$ m. For $z_0 \simeq 10^{-3}$~m, STEP could constrain up to $n\lesssim 1.3$. 

A much stronger constraint is obtained in the context of TeVeS, whose interpolating function tends to $\mu(y)_{TVS}\approx 1-1/y^{1/2}$ for large $y$. This leads in turn to $z_0 /y^{1/2}\gtrsim 10^{-16}$, which is satisfied for $z_0>10^{-10}$~m---surely the case for STEP. Therefore, STEP will easily be capable of falsifying Bekenstein's interpolating function in the context of modified inertia, cross-checking the results from the Cassini spacecraft, which have ruled it out in the scope of modified gravity \cite{2014PhRvD..89j2002H}.

\section{Conclusions and Discussion}

Due to the cumulative nature of its observations as well as its intrinsic sensitivity, STEP constitutes a powerful test of both MOND (and by extension the dark matter paradigm) and the SEP. (Indeed we find that MOND in its modified inertia formulation is inextricably linked to SEP violation, and a test of one must also consider the other.)
In particular, we have shown that any difference in the frequency of oscillation of the STEP test masses relative to Newtonian expectations (but not in their amplitude) would imply the validity of the SEP within MOND.
Conversely, a difference in amplitude but not frequency would imply a violation of the SEP within MOND.
\textit{No} observable difference in frequency \textit{or} amplitude can be interpreted either as a confirmation of Newton's second law and falsification of MOND (if the SEP is valid), or as a constraint on MOND (with SEP violation).
Both the widely-used ``$n$''-family of interpolating functions and the TeVeS formulation of MOND can be constrained or excluded across a significant portion of the theoretical parameter space, even when MONDian effects are ``screened'' by violations of the SEP.

These conclusions are intrinsically related to STEP's state of motion (free fall). Nevertheless, nothing precludes measurements in Earth-based laboratories, for instance, during STEP calibration tests. In this case, though, MOND analyses change because the gravitational acceleration of Earth on the experiment can always be eliminated, thus decreasing $y$. More specifically, now $y\approx |\vec{b}_{c.m}|/a_0\simeq 10^{8}$ (the average norm of the centrifugal acceleration of a particle at rest on Earth, the largest kinematic acceleration now present, is of the order of $10^{-2}$~ms$^{-2}$). In principle one could allow the experiment to run indefinitely, but let us assume it also operates for some days. Given that the position sensitivity is an intrinsic property of STEP \cite{1978AcAau...5...27W}, for the $n$-family interpolating functions this also implies $z_0|n-2| /(ny^n) \gtrsim 10^{-16}$, so that now $n>1$ whenever $z_0>10^{-8}$~m. Larger $n$ could also be investigated if the experiment ran longer, had a higher intrinsic frequency $\omega$ or larger position sensitivity. This latter possibility might be realized in the context of STEP by increasing the current in the SQUID circuits. Such an experiment might even be carried out on the ground before launch, as a ``synergistic'' test of MOND and the SEP (but not, of course, the WEP). By contrast, Earth-based torsion balances, or the recently launched MICROSCOPE mission \cite{2009SSRv..148..455T,2012CQGra..29r4010T}, would not be  ideal for testing Newton's second law since they mainly operate with test particles at equilibrium, where dynamics is suppressed.

\vspace{1mm}

\begin{acknowledgments}
Thanks go to Paul Worden for valuable comments and technical information regarding the details of the STEP experiment. J.P.P. acknowledges the support given by CNPq- Conselho Nacional de Desenvolvimento Cient\'ifico e Tecnol\'ogico of the Brazilian government within the postdoctoral program ``Science without Borders''. A.J.P. thanks the Fisher College of Science and Mathematics and the Department of Physics, Astronomy and Geosciences of Towson University for travel support.
\end{acknowledgments}

\bibliographystyle{apsrev4-1}
\bibliography{MOND_STEP}

\end{document}